# Machine-Learning-Guided CALPHAD Design of Ru-Stabilized BCC–B2 Refractory Alloys

**Nathan Peterson[1], Avik Mahata[2], Nick Beaver[2], Mohsen Kivy[1]**

[1]Department of Materials Engineering, California Polytechnic State University, San Luis Obispo, CA 93407, USA
[2]Department of Mechanical and Electrical Engineering, Merrimack College, North Andover, MA 01845, USA

## ABSTRACT

Refractory alloys with a ductile body-centered-cubic (BCC) matrix strengthened by ordered B2 precipitates offer a high-temperature analogue to the γ/γ′ architecture of Ni-based superalloys. Ruthenium is particularly attractive as a B2 stabilizer because RuHf, RuZr, and RuTi can retain ordered phases well above 1300 °C. In this work, equilibrium CALPHAD calculations were coupled with random-forest-guided active learning to explore a ten-element Nb-based, Ru-bearing composition space containing Nb, Ta, Mo, V, Ru, Ti, Zr, Hf, Al, and Y at 1 at. % resolution. Across 500 CALPHAD-evaluated alloys, the calculations reproduced the principal trends reported for the Ru-B2 design space. RuHf and RuZr remained stable to the solidus, RuTi commonly exhibited a solutionizing window, and Al-containing alloys preferentially formed competing sigma and A15 phases. The upper bound of the BCC+B2 field increased from a median of approximately 1570 °C at 5 at. % Ru to approximately 1980 °C near 9–10 at.% Ru. Among the group-IV additions, Hf, Zr, and Ti produced progressively lower two-phase stability. The dataset was then re-screened using four physically motivated criteria consisting of BCC+B2 stability above 1300 °C, absence of liquid, phase purity, and a secondary-phase fraction between 15 and 55% at 1300 °C. One hundred alloys satisfied all four criteria, including 19 Ru-lean compositions, and two independently reported HfRu-B2 alloys were recovered within this subset. Analysis of the original acquisition function showed that excessive uncertainty weighting and the use of total BCC/B2 fraction biased the search toward poorly ranked compositions. These results establish practical compositional design rules for Ru-stabilized dual phse refractory alloys and identify phase-specific BCC/B2 lattice misfit as the key quantity for the next design cycle.

## 1. Introduction

Ni-based superalloys derive their high-temperature performance from a two-phase γ/γ′ architecture, in which a ductile FCC matrix is strengthened by an ordered precipitate phase [1–5]. This architecture works well, but its useful temperature range is limited by how stable and coherent γ′ remains at high temperature [3,6,7]. Refractory high-entropy and multi-principal-element alloys can reach much higher melting temperatures [1,8–11], but single-phase refractory solid solutions still have not matched the strength, ductility, and long-term microstructural stability of γ/γ′ [2,3].

A BCC matrix strengthened by ordered B2 precipitates is a natural refractory analogue to γ/γ′ [12,13]. Many refractory elements form stable BCC phases, and B2 order offers a route to precipitation strengthening in these types of alloys [14,15]. The main difficulties in designing BCC+B2 microstructures are high temperature stability, the presence of brittle competing intermetallics, and keeping the BCC/B2 interface coherent enough to hold a stable precipitate shape [3,16,17]. Two-phase BCC+B2 refractory alloys have already been demonstrated, but the most studied B2-forming alloying systems lose stability not far above 1200 °C and can turn into a brittle, continuous B2 matrix instead of staying a discrete precipitate [17,18].

Ruthenium stands out as a promising B2-forming addition for refractory alloys. Several Ru-based B2 phases stay stable well above 1300 °C, while conventional refractory pairings tend to order weakly, if at all [3,19]. A first-principles survey of B2 formation across the refractory metals reaches a consistent conclusion on why. Ordinary group IV–VI pairings order weakly, and stable ordering instead requires a group IV/V element paired with a group VII/VIII element such as rhenium, ruthenium, or osmium [2,20]. This is the mechanistic reason Ru is used here as the primary B2 stabilizer rather than a more conventional refractory pairing. However, even with a somewhat narrower design space due to the focus on Ru-B2 alloys in this work, the composition space is still too large to search exhaustively with CALPHAD alone. In this work, we use a fast surrogate model to rank a broad candidate pool and save the expensive equilibrium calculations for the candidates most worth calculating [21–24].

The focus on an Nb matrix is deliberate. Niobium is a well-established refractory base element that has already been explored alongside Ta, Mo, and V in related Ru–B2 studies [3,8,25]. Including the additional elements in the design space allows the search to test multiple routes to B2 formation and shift the balance between matrix stability and precipitation behavior [26]. The result is a design space that stays realistic but still bounded, so the model learns from comparable compositions instead of an unrestricted, chemically incoherent list of alloys.

Kube and co-workers evaluated roughly 3500 ternary and quaternary systems formed from 24 practical B2 phases and 153 matrix-element combinations, calculating phase equilibria on composition grids spaced at 10 at. % and subsequently validating 20 selected alloys experimentally. [3]. Follow-on studies from the same group measured B2 solvus lines, misfits, and grain-boundary phases in HfRu- and ZrRu-bearing Nb alloys [27], demonstrated a solution-and-age pathway for HfRu-B2 [28], and showed that Ru contents above roughly 9 at.% promote cracking under rapid solidification [29] [30]. Two gaps remain. The pseudobinary grid does not sample the multi-element interior of the space, where four to six elements compete simultaneously, and lattice misfit enters those studies as a measurement made after the alloys were chosen rather than as a term in the selection itself. This work addresses both. The search runs at 1 at. % resolution over four to six elements drawn from a ten-element pool, and a compositional mismatch penalty sits inside the acquisition function that decides which alloys are calculated at all.

A BCC/B2 pairing can be nominally optimal for B2 formation and still make a poor precipitation-strengthened material if the phase fractions, matrix chemistry, or interfacial mismatch are wrong [18,31,32]. Therefore, this work is organized around four CALPHAD-based design goals rather than a single pass/fail criterion. First, the upper bound of the two-variant BCC + B2 field must exceed 1300 °C. Second, no liquid may be present at or below 1300 °C. Third, the equilibrium constitution must contain only BCC_B2-family phases, since brittle or complex competing intermetallics, including Laves, sigma, and A15-type phases, can ruin an otherwise favorable design even when the stability target is met [33,34]. Fourth, the equilibrium constitution must contain a precipitate, which the other criteria do not require, so a precipitate phase fraction between 15 and 55% at 1300 °C is applied as a separate condition. Alongside these four CALPHAD screening criteria, a compositional atomic-size mismatch descriptor δ [35], calculated from the bulk composition, is retained in the active-learning objective as a fast proxy for lattice compatibility rather than a substitute for the phase-specific BCC/B2 lattice-parameter mismatch that ultimately controls coherency [6,36,37].

In this work, we combined equilibrium CALPHAD calculations with a random-forest-guided active-learning framework to search a ten-element, Nb-based, Ru-bearing composition space at 1 at. % resolution. The workflow was used to rank candidate alloys, identify the effects of individual alloying elements on BCC+B2 stability, phase purity, and precipitate formation, and assess the behavior of the acquisition function itself. From 500 CALPHAD-evaluated compositions, 100 alloys satisfied all four screening criteria and 19 remained within the Ru-lean fabrication range, with the most promising candidates concentrated in Nb-Ta-Ru-Hf and Nb-Mo-Ru-Hf chemistries. Two of the selected compositions also coincide with alloys reported independently in experimental literature, providing an external check on the search strategy.

# 2. METHODS

## 2.1 CALPHAD Calculation Validation

Every candidate composition was evaluated with an equilibrium calculation (Figure 1) in Thermo-Calc using the TCHEA4 database, sweeping temperature from 1000 to 2500 ℃ in 10 ℃ increments to compute the stable phase constitution at each step. The phase identities, fractions, and the resulting BCC_B2 solvus and solidus were used to assess the alloy based on th ‘kmye optimization criteria above. Only the initial 20 seed alloys were checked by hand, confirming that each equilibrium step diagram behaved physically sensibly before the active-learning loop was allowed to run. Calculations that failed to converge were dropped rather than labeled, since a non-converged run does not distinguish a genuinely poor alloy from a numerically difficult one. After validation, the active-learning loop was run until it reached 500 validated calculations, an arbitrary stopping point rather than a targeted sample size. Figure 1 shows four alloys drawn from the search, chosen to span the range of behaviour the calculations produced. Section 3.2 lists the alloys that pass all four screening goals. Two temperature measures are reported from these step diagrams and they are kept distinct throughout. The first is the upper bound of the BCC_B2 family, the highest temperature at which any BCC_B2-labelled phase remains above the 0.5% threshold, which is the quantity used for the thermal-stability screen in Section 3.1. The second is the upper bound of the two-variant BCC + B2 field, the highest temperature at which at least two distinct BCC_B2 variants coexist, which is the quantity that corresponds to a genuine matrix and precipitate architecture and is used for the element-level comparisons in Section 3.3.

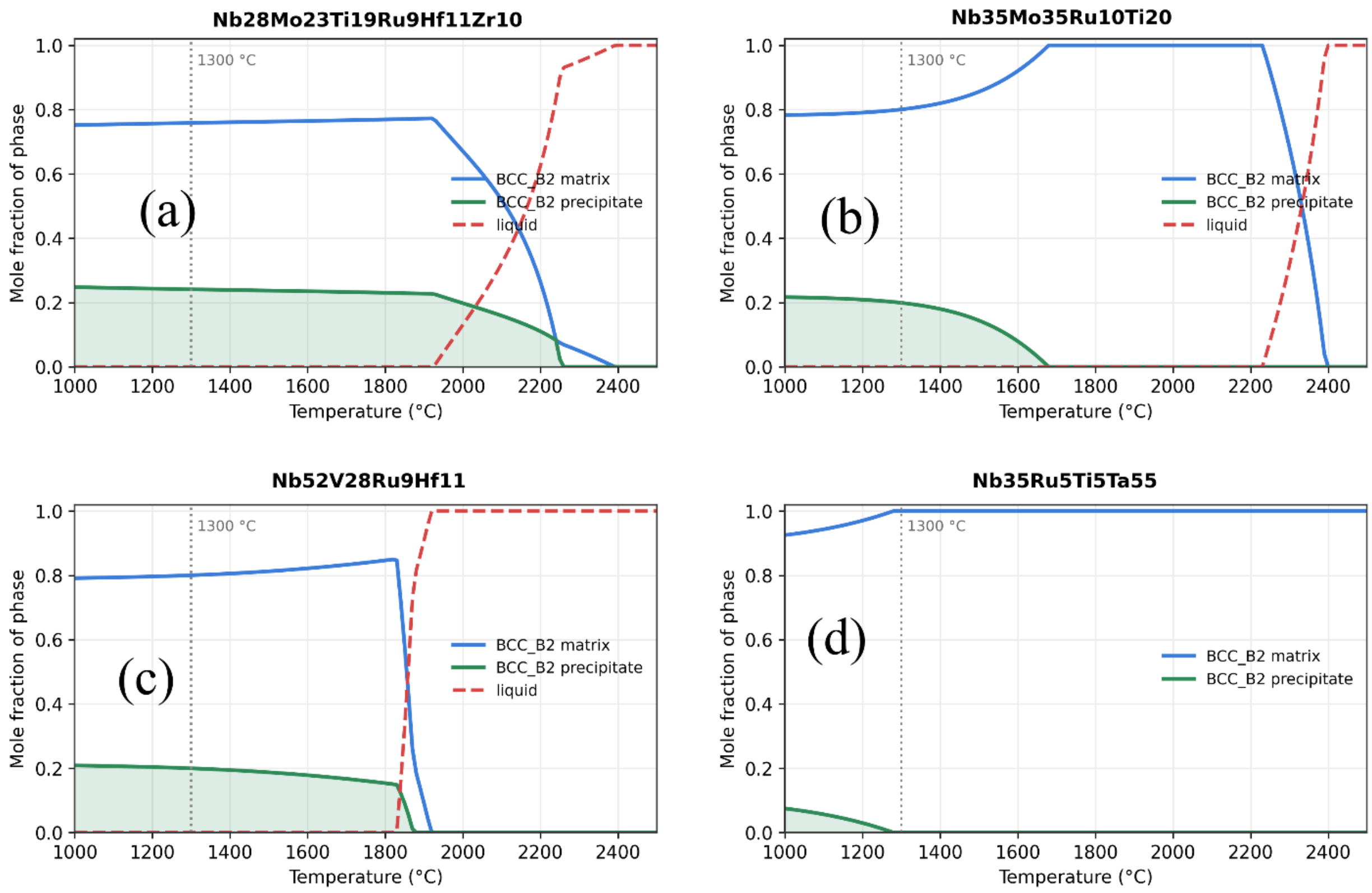

**Figure 1.** Four representative equilibrium step diagrams spanning the behaviour found across the 500 alloys, replotted from the raw Thermo-Calc output. (a) Nb28Mo23Ti19Ru9Hf11Zr10, the best alloy of the passing set, with a two-phase field to 2250 °C and a 24% precipitate fraction. (b) Nb35Mo35Ru10Ti20, a RuTi alloy whose solvus lies 570 °C below the solidus and therefore offers a dissolution and ageing window. (c) Nb52V28Ru9Hf11, a RuHf alloy in which the B2 persists to melt with no solvus. (d) Nb35Ru5Ti5Ta55, which clears the stability and purity criteria yet is single phase above 1300 °C and forms no precipitate at all.

## 2.2 Design Space and Alloy Selection

Candidate compositions were drawn from Nb, Ta, Mo, V, Ti, Ru, Hf, Zr, Al, and Y. Every candidate contained at least 20 at. % Nb and 5 at. % Ru, no single element exceeded 65 at.%, and each candidate carried four to six elements at 1 at. % resolution. Sampling was matrix biased. Most candidates paired Nb with Ta, Mo, or V as a second matrix-forming element, with remaining slots filled from the rest of the pool. This kept the search anchored to experimentally relevant Nb-based matrix chemistries while still letting the other elements compete as B2-forming or modifying additions. These bounds were fixed before the equilibrium calculations and active-learning loop were run. The single-element cap prevents any one element from dominating a candidate, and the limited number of elements per composition keeps each alloy compositionally complex but still interpretable. None of these rules guarantee a two-phase microstructure on their own. They exist only to create a sensible design space consisting of an Nb-dominant matrix with Ru additions for B2 stability, so downstream CALPHAD results stay comparable to one another.

### 2.3 Composition descriptors

Each alloy was represented by seven rule-of-mixtures composition descriptors, summarized in Table 1: mean atomic radius, mean melting point, valence-electron concentration, mixing enthalpy, mixing entropy, atomic-size mismatch, and the Ω parameter. These seven were retained because they carried the strongest relationship to the CALPHAD-derived targets among the descriptors evaluated, not primarily for simplicity [31]. The full feature vector passed to each random forest also includes the ten raw elemental fractions alongside these seven descriptors. The mismatch descriptor δ is also one of the four properties a random forest is trained to predict as described in Section 2.4.

**Table 1.** The seven composition descriptors used to represent each candidate alloy.

| Descriptor | Formula | Description |
|---|---|---|
| Mean atomic radius | $\bar{r} = \sum_i c_i r_i$ | Composition-weighted average atomic radius of the constituent elements |
| Mean melting point | $\bar{T}_m = \sum_i c_i T_{m,i}$ | Composition-weighted average elemental melting temperature |
| Valence-electron concentration (VEC) [32] | $\mathrm{VEC} = \sum_i c_i \mathrm{VEC}_i$ | Composition-weighted average number of valence electrons per atom |
| Mixing enthalpy (ΔHmix) [33–35] | $\Delta H_{\mathrm{mix}} = 4 \sum_{i<j} \Omega_{ij} c_i c_j$ | Regular-solution estimate, summed over binary element pairs weighted by composition |
| Mixing entropy (ΔSmix) [33,34] | $\Delta S_{\mathrm{mix}} = -R \sum_i c_i \ln c_i$ | Ideal configurational entropy of mixing ($R = 8.314\ \mathrm{J\ mol^{-1}\ K^{-1}}$) |
| Atomic-size mismatch (δ) [36] | $\delta = 100 \sqrt{\sum_i c_i \left(1 - \frac{r_i}{\bar{r}}\right)^2}$ | Hume-Rothery-style spread in atomic radii relative to the mean |
| Ω parameter [37] | $\Omega = \frac{\bar{T}_m \cdot \Delta S_{\mathrm{mix}}}{\mid \Delta H_{\mathrm{mix}} \mid}$ | Combines melting temperature, mixing entropy, and mixing enthalpy into a single solid-solution indicator |

### 2.4 Random forests and the active-learning loop

A random forest is an ensemble of decision trees, each trained on a bootstrapped sample of the data; averaging their individual predictions gives a single estimate, and the spread across trees gives a practical measure of how confident the model is in that estimate [40], [41]. Random forests suit our optimization problem well because the training set is small and tabular, and because the tree-to-tree spread gives a usable per-prediction uncertainty estimate [42], [43].

Four random forests were trained on each generation's accumulated dataset, one for BCC-B2 solvus temperature, one for the compositional mismatch δ, one for peak BCC_B2 fraction, and one for phase count. Each forest takes the same input vector, the ten raw elemental fractions plus the seven descriptors from Table 1, and was implemented with scikit-learn using 300 trees per forest [38]. Training a forest on δ is somewhat redundant, since δ can also be computed directly from composition, but it lets the acquisition function reuse the same ensemble-uncertainty machinery for δ that it uses for the other three targets.

Each generation sampled 10,000 untested compositions and scored them with an upper-confidence-bound (UCB) acquisition function. Predicted solvus and predicted BCC_B2 fraction were rewarded, along with their ensemble spread, to favor both high performance and high model uncertainty; predicted phase count was penalized, offset by its own spread; and the exact compositional δ was penalized, offset by the δ-model's ensemble spread:

$$Score = 0.20(\hat{y}_{solvus} + k.\sigma_{solvus}) + 0.25(\hat{y}_{BCC} + k.\sigma_{BCC}) - 0.40(\hat{y}_{phases} - k.\sigma_{phases}) - 0.15(\delta_{exact} - k\sigma_{\delta})$$

where ŷ is a forest's min–max normalized mean prediction, σ its normalized ensemble spread, δexact the exact compositional mismatch (not the forest's prediction of it), and κ = 3 the exploration weight. Weighting phase count at 40%, BCC_B2 fraction at 25%, solvus at 20%, and δ at 15% means avoiding competing phases dominated the ranking, and the compositional-mismatch penalty played a comparatively small role in which candidates were chosen, a point revisited in Section 3.5. The highest-scoring candidates from each generation were sent to Thermo-Calc, added to the training set, and used to retrain all four forests for the next generation, following the same exploit and explore logic as UCB methods in the Bayesian-optimization literature [39,40]. Two properties of this score govern its behaviour and are examined in Section 3.7. The mean and the spread are each min-max normalized over their own range before κ multiplies the spread. Because σ is far smaller than the spread of ŷ in physical units, normalizing them separately inflates σ to the same 0 to 1 range, so κ = 3 makes every exploration term three times its exploitation term. Expanding the phase-count contribution gives $-0.40 \cdot \hat{y}_{phases} + 1.20 \cdot \sigma_{phases}$, and because tree disagreement is largest where alloys are compositionally messy, the uncertainty bonus outweighs the penalty it was meant to apply. Separately, δ enters the score as an exact value while the other three terms are noisy predictions, and δ is also one of the 17 input features, so the δ forest reproduces it with $R^2 = 1.000$ and contributes no exploration bonus at all.

**Figure 2.** (a) Random-forest and active-learning workflow used for CALPHAD-guided alloy screening. (a) A random forest generates the mean prediction and ensemble-based uncertainty from 300 bootstrapped decision trees. (b) Four random-forest models rank candidate compositions using the acquisition function, after which the highest-scoring alloys are evaluated by equilibrium CALPHAD calculations and returned to the training set for the next generation.

### 2.5 Data processing and label construction

Two thresholding rules turned raw Thermo-Calc output into training labels. A phase was counted as present, for both the BCC_B2 solvus and the phase-count target, only once its fraction exceeded 0.5%, which filters out trace numerical noise while keeping anything large enough to matter to the screening decision. Symmetry-split BCC/B2 variants (BCC/B2, BCC_B2#2, BCC/B2#3, and so on) were summed and treated as a single phase family, since the database can report compositionally distinct regions of the same ordered/disordered pair separately even when they represent one matrix/precipitate architecture. Applying these rules to the converged calculations produced the dataset used throughout the Results below. A third quantity is used throughout the Results. At each temperature the BCC_B2 variants were sorted by magnitude rather than by their composition-set label, since Thermo-Calc interchanges the #1, #2, and #3 labels between temperature steps. The largest variant is taken as the matrix and the sum of the remainder as the precipitate, so the precipitate phase fraction at 1300 °C follows directly from the step diagram. Two temperature measures are also kept distinct. The upper bound of the BCC_B2 family is the highest temperature at which any BCC_B2-labelled phase exceeds the threshold, which is the quantity used for the original thermal-stability screen. The upper bound of the two-variant BCC + B2 field is the highest temperature at which two or more distinct variants coexist, which is the quantity that corresponds to a matrix and precipitate architecture and is used everywhere below.

Reading the phase labels also requires care, because names such as CR3SI_A15 and AL2ZR3 are structure prototypes in the TCHEA database rather than formulas. All 26 alloys reported to contain CR3SI_A15 are free of both Cr and Si, so these labels are read here as A15 and as prototype-named intermetallics.

## 3. Results and Discussions

The active-learning search evaluated 500 CALPHAD-validated alloys against the four CALPHAD screening criteria set out in the Introduction. Section 3.1 quantifies how often each criterion is met and then applies the final four-criterion screen. Section 3.2 examines the alloys that satisfy all four conditions. Section 3.3 reads the same dataset element by element, which turns the pass rates into design guidance, and Sections 3.4 to 3.6 return to the search behaviour and to the mismatch descriptor that shapes the broader design space. The precipitate phase-fraction criterion behaves differently from the other three because it separates a true BCC + B2 superalloy constitution from a single-phase alloy. Section 3.7 then examines the screening layer itself.

### 3.1 Attainment of the four screening goals

Table 2 summarizes how often each screening criterion was met across the 500 CALPHAD-validated alloys. Evaluated on the final four CALPHAD-based criteria, thermal stability is met by 389 alloys (77.8%), absence of liquid at or below 1300 °C by 296 (59.2%), phase purity by 173 (34.6%), and a precipitate phase fraction in the useful 15 to 55% window by 307 (61.4%). One hundred alloys, or 20.0% of the dataset, satisfy all four simultaneously. The compositional mismatch descriptor $\delta$ is reported separately for reference. Only 13 alloys (2.6%) fall below $\delta < 2\%$, which shows that low bulk mismatch is comparatively rare in this design space. The two compositions Nb20Ru5Ti30Ta45 and Nb35Ru5Ti5Ta55 satisfy the stability, no-liquid, and phase-purity criteria but do not satisfy the precipitate criterion. Their precipitate phase fractions at 1300 °C are 0.05 and 0.00, respectively, so neither realizes the two-phase constitution targeted here.

**Table 2.** Design-goal attainment across the 500 CALPHAD-validated alloys. Thermal stability is measured by the upper bound of the two-variant BCC + B2 field. The compositional descriptor $\delta$ is reported for reference and is not part of the final four-criterion down-selection.

| Design goal | Screening criterion | Passing alloys | Pass rate |
|---|---|---|---|
| Thermal stability | Two-variant BCC + B2 field > 1300 °C | 389 / 500 | 77.8% |
| No incipient melting | No liquid at or below 1300 °C | 296 / 500 | 59.2% |
| Phase purity | Only BCC_B2-family phases (liquid excluded) | 173 / 500 | 34.6% |
| Low mismatch | Compositional descriptor $\delta < 2\%$, see Section 3.6 | 13 / 500 | 2.6% |
| Precipitate fraction | $15\% \leq$ precipitate fraction at 1300 °C $\leq 55\%$ | 307 / 500 | 61.4% |
| All four goals | Stability, no melting, phase purity and precipitate fraction | 100 / 500 | 20.0% |

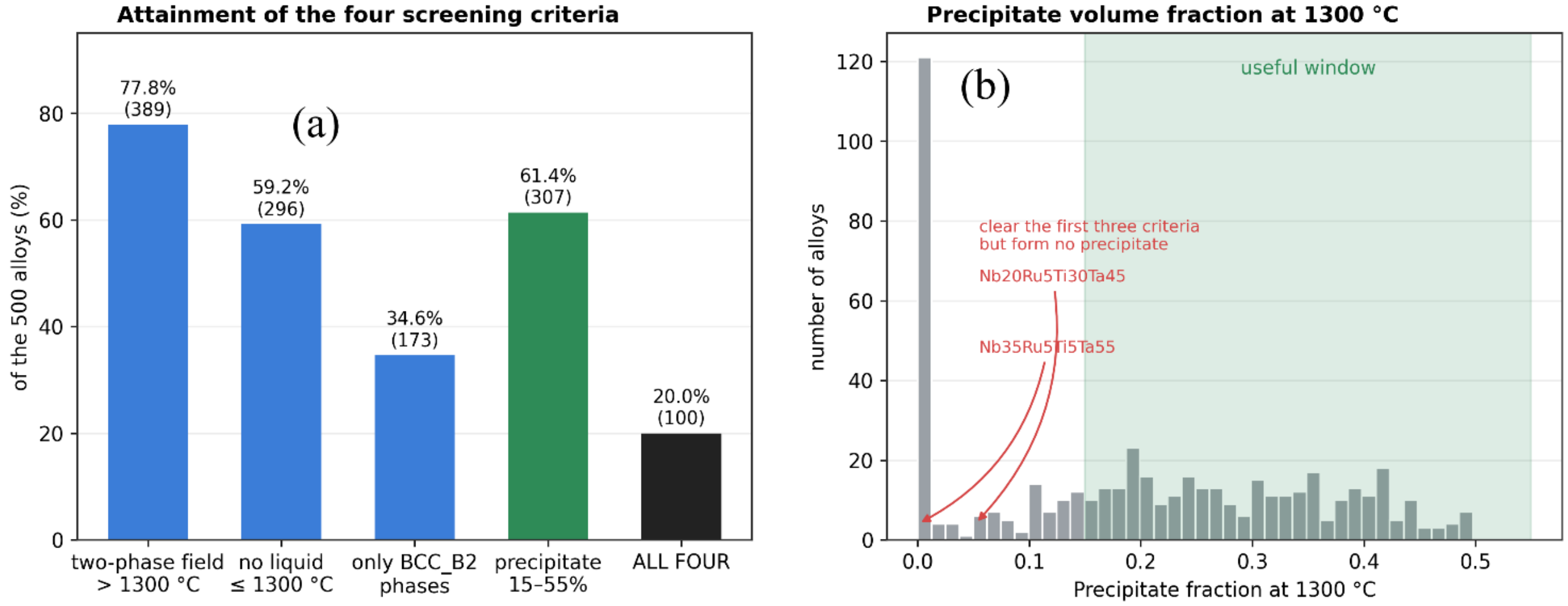

**Figure 3.** Screening outcome across the 500 alloys. (A) Attainment of each of the four CALPHAD screening criteria. (B) Distribution of the precipitate phase fraction at 1300 °C, with the useful window shaded and the two alloys that satisfy the stability, no-liquid, and phase-purity criteria without forming a useful precipitate marked.

### 3.2 The Alloys That Pass all Four Criteria

Of the 100 alloys that satisfy every criterion, 19 also hold Ru at or below 10 at.%, the level above which single-track laser melting of Ru-bearing BCC-B2 alloys shows cracking on cooling [29,30]. That subset is the practical down-selection from this dataset. The highest two-phase-field member is Nb28Mo23Ti19Ru9Hf11Zr10, which holds a two-variant BCC + B2 field to 2250 °C with a solidus of 1930 °C, a precipitate fraction of 0.24, and no competing phases at 9 at.% Ru. Its compositional δ is 5.84%, so a strict δ screen at the 2% level excludes it. Nb50Mo30Ru10Hf10, Nb52Ta28Ru9Hf11, and Nb55Ta20Ru10Hf15 follow at fields of 2120, 2070, and 2050 °C with precipitate fractions of 0.20 to 0.25. Two entries in that list are independent checks rather than predictions. Nb52Mo28Ru9Hf11 and Nb52V28Ru9Hf11 are the compositions reported experimentally for solution and aged HfRu-B2 alloys [28], and the present calculations place them at fields of 2110 and 1870 °C with precipitate fractions of 0.20 and no competing phases, which puts both in the top half of the fabricable passing set. The screen therefore recovers alloys that were arrived at experimentally by a different route, which is the strongest available evidence that the underlying equilibrium calculations are sound.

The two compositions that satisfy the stability, no-liquid, and phase-purity criteria are Nb20Ru5Ti30Ta45 and Nb35Ru5Ti5Ta55. Nb20Ru5Ti30Ta45 exhibits lower Nb content offset by higher Ti, whereas Nb35Ru5Ti5Ta55 is more Nb-Ta-rich. Both alloys diverge from the Al/Ti-rich compositional trend the active-learning loop favored on average (Section 3.5). Comparative characterization of phase fractions and elemental partitioning following identical thermal treatment would establish whether the two compositions attain the target BCC + B2 balance via a common mechanism or through distinct partitioning pathways, and the broader 500-alloy dataset can inform incremental compositional refinement of either candidate as needed. Both candidates also sit on the fabricable side of a known processing limit. Single-track laser melting of Ru-bearing BCC-B2 alloys shows crack formation on cooling once Ru exceeds roughly 9 at.%, and solidification cracking in Zr-bearing variants is mitigated by substituting Hf [29,30]. Holding Ru at the 5 at.% floor therefore costs roughly 400 °C of two-phase field relative to the 9 to 10 at.% optimum identified in Section 3.3 and buys back a composition that can plausibly be built. Neither develops a useful precipitate, however, so neither can deliver the strengthening architecture the search targets, and neither is carried forward. They are useful instead as worked examples of what a screen without a precipitate criterion selects.

### 3.3 Element-Level Trends Behind the Screening Outcomes

The pass rates in Table 2 say how often the design goals were met but not which chemistries were responsible. Reading the same 500 step diagrams element by element supplies that, and it recovers a set of trends that can be stated without any further calculation. Ruthenium sets the height of the two-phase field (Figure 4A). The median upper bound of the two-variant BCC + B2 field is about 1570 °C in alloys holding Ru at the 5 at.% floor, rises to about 1980 °C at 9 to 10 at.% Ru, and then flattens out, so the return on additional Ru is spent by roughly 10 at.%. Among the group-IV additions the ordering is unambiguous (Figure 4B). Taking only those alloys that carry a single group-IV element, the median two-variant bound is 2000 °C for Hf, 1765 °C for Zr, and 1480 °C for Ti. This reproduces, from equilibrium thermodynamics alone, the experimental ranking of HfRu above ZrRu measured by sequential annealing of Nb-based alloys [27]. Phase purity separates the matrix-forming additions from the functional ones (Figure 4C). Vanadium is the only element in the pool that raises purity substantially, from 27% when absent to 49% when present, and it also raises the median solidus, although it lowers the two-variant bound. Tantalum raises the two-variant bound from 1770 to 1930 °C and the median solidus from 1380 to 1470 °C, and the only two alloys in the dataset with no liquid anywhere below 2500 °C, Nb20Ru5Ta60V15 and Nb35Ru5Ti5Ta55, are both Ta-rich at low Ru. Yttrium and aluminium are the two clear liabilities. Purity falls to 12% in Y-bearing alloys against 53% without Y, and 81% of Y-bearing alloys carry liquid at or below 1300 °C with a median solidus of 1180 °C, so yttrium is disqualifying in this design space rather than merely unhelpful. Aluminium lowers purity from 43% to 26%.

The competing phases themselves trace cleanly back to individual additions (Figure 4D), which is the most directly usable result in this section. Sigma and A15 occur in none of the 255 Al-free alloys and in 106 and 26 of the 245 Al-bearing ones, respectively, and sigma forms in 77 of the 83 alloys that carry both Al and Ta. Aluminium is therefore the sole driver of the two topologically close-packed phases that most threaten toughness here, and its damage is concentrated where it meets Ta. That conclusion matches the experimental survey of the Ru-B2 space, in which RuAl alloys solutionize when combined with V and Mo but form competing sigma and A15 phases when combined with Nb and Ta [3]. HCP phases are almost entirely a Y effect, appearing in 74 of the 77 cases alongside Y, and the Ru-Y intermetallics RuY3-D011 and Ru2Y5 account for a further 110 occurrences. Laves phases, by contrast, are rare, with 18 C14 and 13 C15 occurrences across the whole dataset, so the phase-purity failures in this space are a sigma, A15, and Ru-Y problem rather than a Laves problem. The practical reading is that Ru fixes the height of the two-phase field, Hf and Ta hold it up, V protects phase purity, and Al and Y are the additions to remove first.

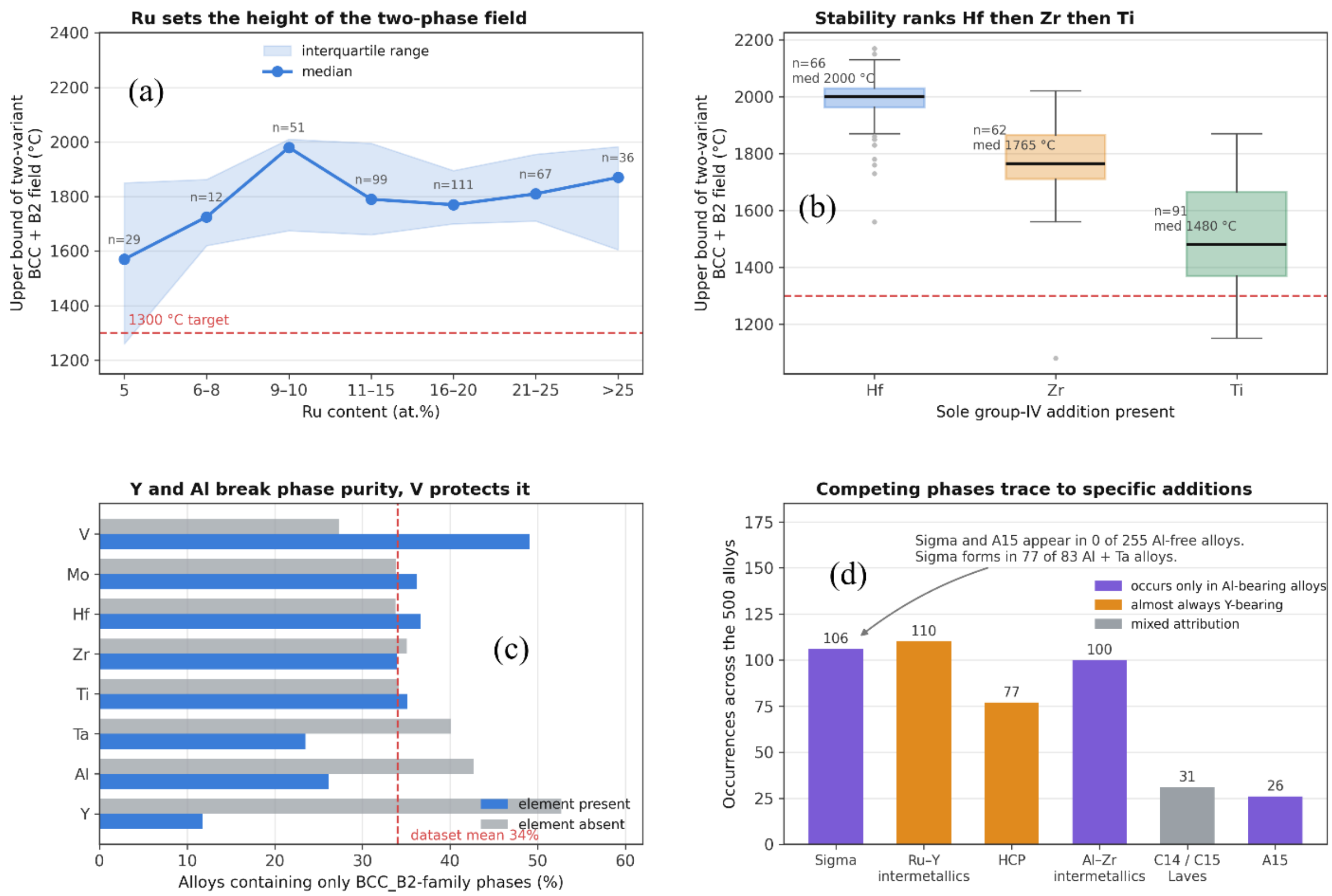

**Figure 4.** Element-level trends across the 500 CALPHAD-validated alloys. (A) Median upper bound of the two-variant BCC + B2 field against Ru content, with the interquartile range shaded. (B) The same bound for alloys carrying a single group-IV addition, ranking Hf above Zr above Ti. (C) Fraction of alloys containing only BCC_B2-family phases, with and without each element. (D) Inventory of competing phases, colour-coded by the addition they track.

### 3.4 Connecting Back to The Design Rationale

These results follow directly from the mechanistic argument and the screening structure set out in the Introduction and Methods. The Ru-driven B2 stabilization mechanism, in which a group IV/V element is paired with Ru, is doing exactly the job it was chosen for. High-temperature stability is attainable across a majority of the design space, which is what the group IV/V mechanism predicts. What the mechanism does not guarantee, and what the four-criterion structure is built to separate out, is phase cleanliness, useful precipitate formation, and coherency. The relative pass rates make the sequential character of the screen explicit. A high solvus is necessary but only weakly selective within this design space. Phase purity is a substantially stronger filter, particularly where alloying additions stabilize phases outside the desired BCC_B2 family. The precipitate criterion is the step that distinguishes a true BCC + B2 architecture from a single-phase alloy. This filtering pattern also explains why Nb20Ru5Ti30Ta45 and Nb35Ru5Ti5Ta55 survive the stability, no-liquid, and phase-purity criteria yet are not carried forward as final candidates. Section 3.3 supplies the mechanism behind each of these behaviours. The solvus is weakly selective because Ru alone carries it above the target once its content passes about 9 at.%. Phase purity is a strong filter because two specific additions, Al and Y, generate almost all of the competing phases. The precipitate criterion is selective because a stable and clean BCC_B2 family does not necessarily partition into matrix and precipitate. The compositional mismatch descriptor remains restrictive because it is governed by atomic-size contrast, which the elements that raise the solvus do not automatically reduce.

### 3.5 Search Trajectory and Compositional Drift

The compositions submitted to CALPHAD changed systematically across the 50 chronological batches used to track the search. Early evaluations skewed more Nb-, Ta-, and Mo-rich; as the loop accumulated data, it increasingly picked compositions containing Al, Ti, Ru, and Zr, while Ta and Mo became less prominent in the typical candidate. This directed shift shows that the ranking changed as CALPHAD results accumulated, but it does not, by itself, show that the workflow outperformed a budget-matched random screen; that comparison was not run here, so the trend should be read as a search trajectory rather than a direct measure of search efficiency. It is also important to keep this trend and the all-goal candidates conceptually separate. The batch-to-batch drift shows where the active-learning loop spent its calculations over time, and is useful for diagnosing how the model behaved, but it does not by itself prove that every Al- or Ti-rich alloy will perform well. In fact the Ru-lean alloys that pass all four goals are aluminium-free, and they occupy a Ta-rich and Hf-rich region rather than the Al-rich region the drift trend would suggest. Section 3.3 explains why that drift was only partly productive. The loop moved towards Al and Zr, and both additions carry the competing phases that the phase-count term was meant to penalize, so a share of the later calculations was spent in a region the model had learned to score well on solvus and BCC_B2 fraction while paying for it on constitution. Section 3.7 shows that this drift is not a record of the model learning. It is the signature of an acquisition function whose exploration terms dominate its exploitation terms, which rewards precisely the compositionally complex, Al-bearing alloys on which the phase-count forest is least certain.

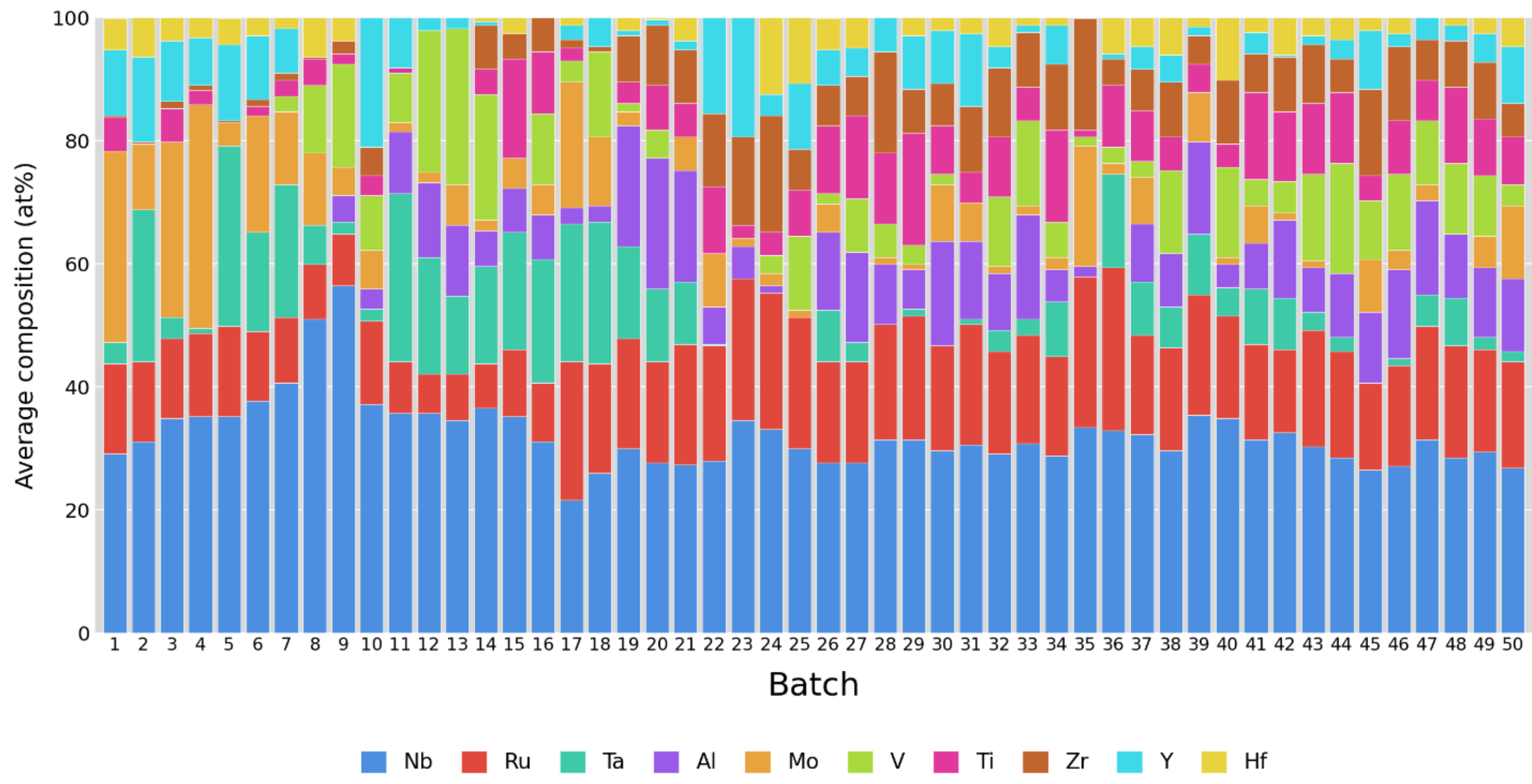

**Figure 5.** Search trajectory across the 50 chronological batches used in the active-learning loop. Each bar gives the mean elemental composition of one batch, corresponding to approximately 10 alloys ordered by file modification date.

### 3.6 Lattice Mismatch as The Dominant Constraint

The δ distribution shows that the large majority of candidates sit well above the 2% screening threshold, while only a handful clear it. This descriptor is calculated from the bulk composition and is used here as a low-cost proxy for lattice compatibility within the active-learning workflow. It is therefore most useful as a comparative screening variable rather than as a direct substitute for the phase-specific BCC/B2 lattice mismatch that ultimately controls coherency [6,36,37]. The scarcity of alloys below $\delta = 2\%$ shows that low bulk mismatch is restrictive within the present ten-element design space. The δ parameter carried only 15% of the acquisition weight (Section 2.4), so the ranking was influenced more strongly by the stability and constitution terms than by mismatch. The distribution also explains why experimentally attractive Hf- and Zr-bearing chemistries tend to reside above the 2% line when δ is computed from the bulk composition. The Goldschmidt radii of Nb, Ta, and Al agree to within 0.2%, so the lowest values cluster in the Nb-Ta-Al-Ti corner and near the 5 at. % Ru floor. Hf-bearing and Zr-bearing alloys therefore rarely satisfy a strict $\delta < 2\%$ cutoff even though several of them perform well on the CALPHAD screening criteria and align with experimentally reported Ru-B2 alloy families [27,28]. In this design space, δ is best interpreted as a useful composition-level proxy that complements a later phase-specific assessment of BCC/B2 misfit. A first-principles framework developed by Mahata et. al. [41] for Ru-containing BCC–B2 alloys has shown that thermodynamic B2 stability and phase-specific lattice coherency must be considered jointly, with the latter evaluated directly from the BCC and B2 lattice parameters. The present study uses the simpler composition-level descriptor during high-throughput screening and reserves the phase-specific treatment for subsequent refinement.

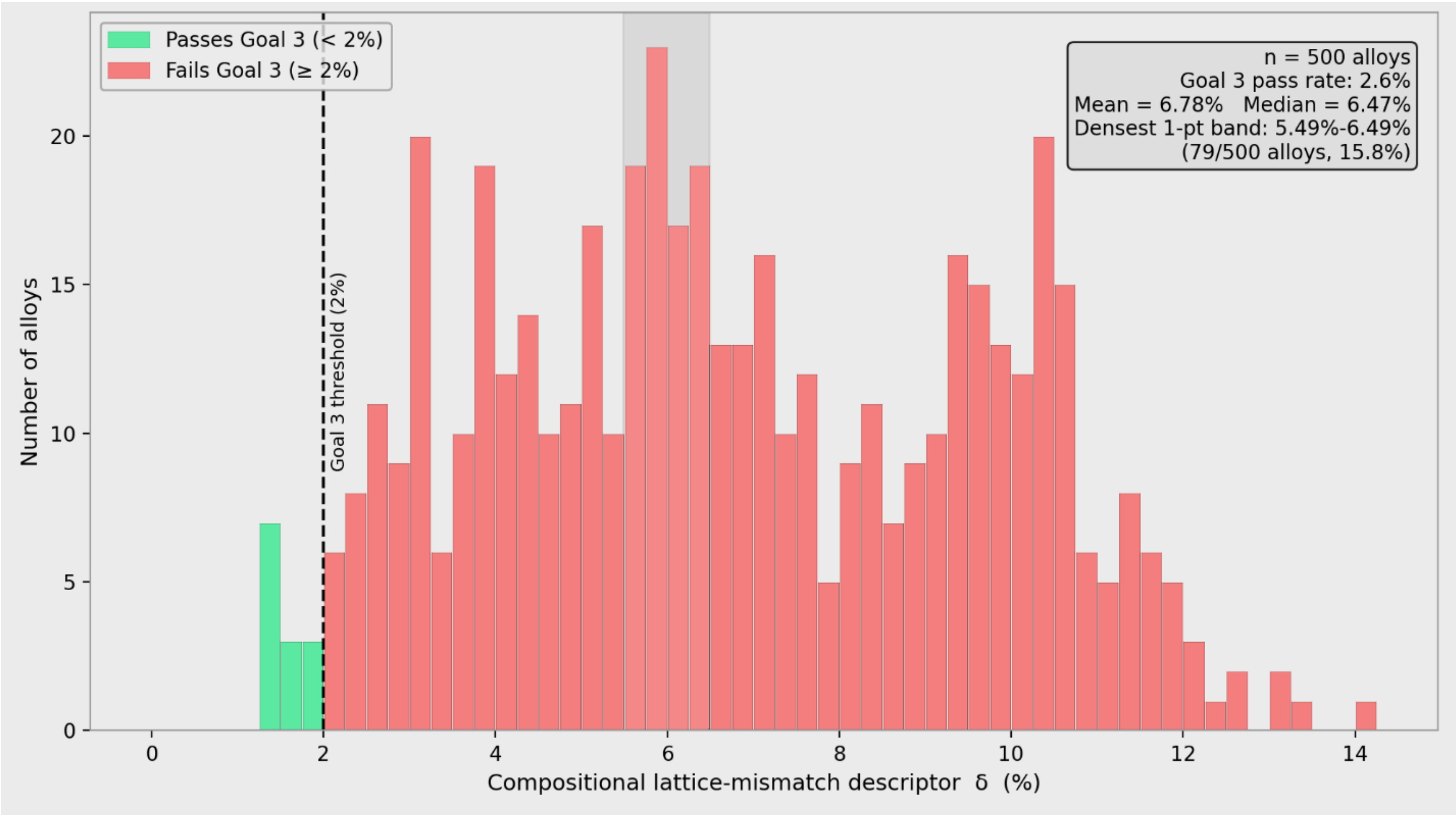

**Figure 6.** Distribution of the compositional lattice-mismatch descriptor δ across the 500 CALPHAD-validated alloys. The dashed line marks the 2% threshold used in the original screening logic.

### 3.7 Model Performance and Audit of the Screening Layer

The four random forests were re-trained on the measured CALPHAD labels and assessed out of fold with five-fold cross-validation (Figure 7). For this post-hoc audit, the targets were the two-variant BCC + B2 field, solidus, precipitate fraction at 1300 °C, and phase count. The two-variant BCC + B2 field is predicted with $R^2 = 0.797$ and a mean absolute error of 68 °C, the solidus with $R^2 = 0.886$ and 82 °C, the precipitate fraction with $R^2 = 0.685$, and the phase count with $R^2 = 0.691$. These are usable accuracies for a dataset of this size, so the surrogate model itself was not the weak element of the workflow.

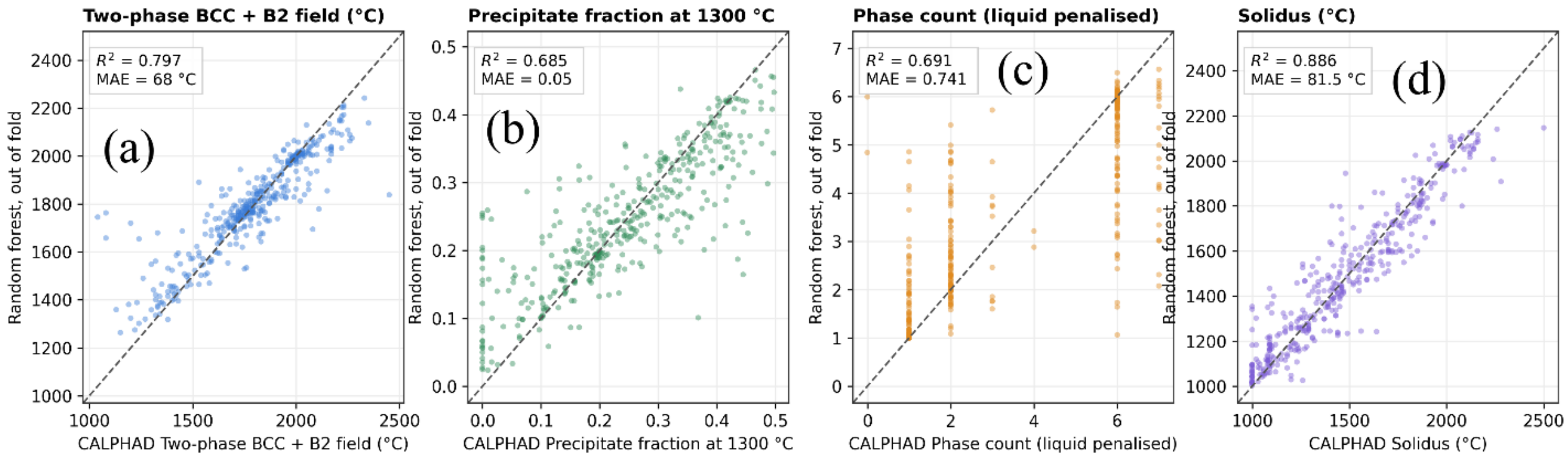

**Figure 7.** Out-of-fold random-forest predictions for the 500 CALPHAD-evaluated alloys. (a) Two-phase BCC+B2 field temperature. (b) Precipitate fraction at 1300 °C. (c) Liquid-penalized phase count. (d) Solidus temperature. Dashed lines indicate perfect agreement between CALPHAD values and random-forest predictions.

Shapley additive explanations resolve what the forests use (Figure 8). Zr, the mean melting temperature, and δ dominate the two-phase field, with Hf and V next, which is the same ordering the element-level analysis of Section 3.3 produced from the raw phase diagrams. For the precipitate fraction the ranking is Zr, Ru, and the mixing enthalpy, so the model attributes precipitate formation to the B2-forming pair rather than to the matrix, consistent with the pseudobinary picture used in the reference screen [3].

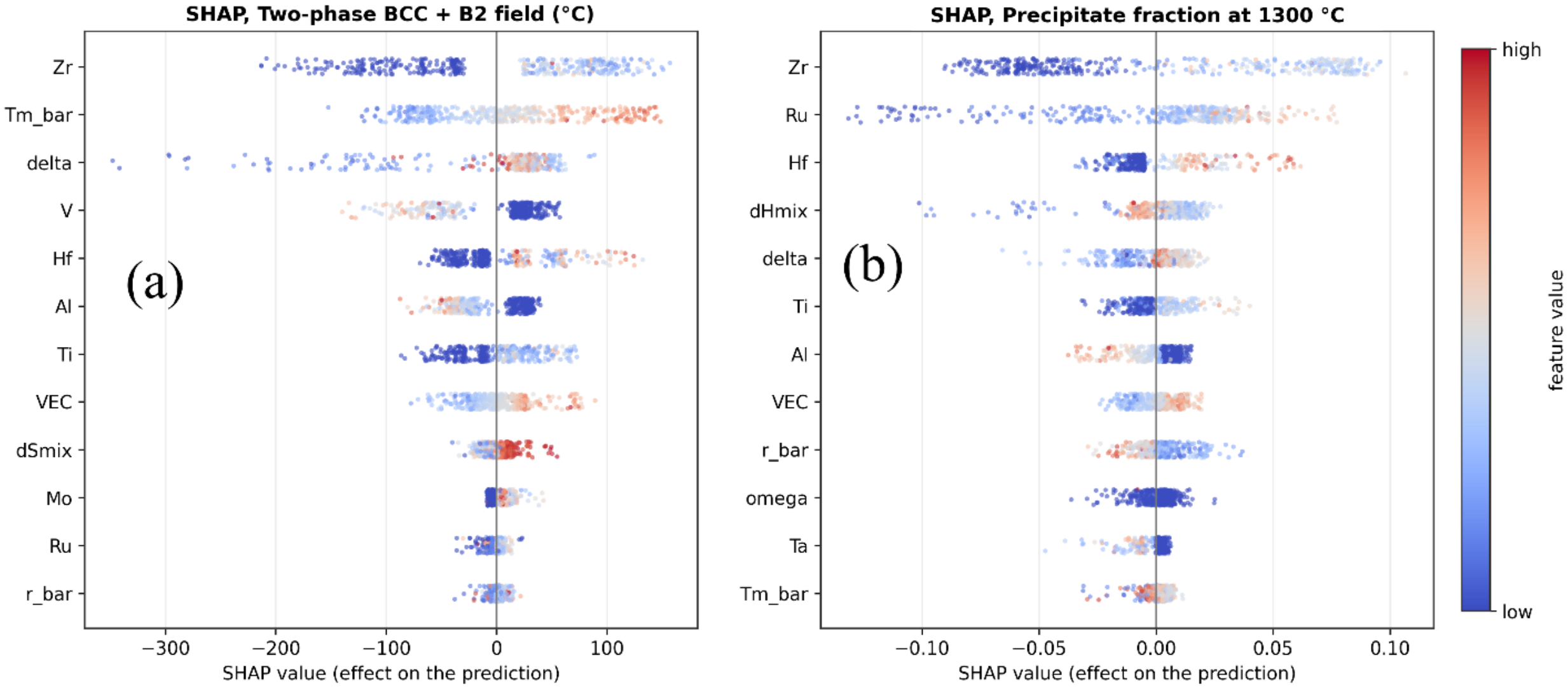

**Figure 8.** SHAP analysis of the random-forest models. (a) Feature contributions to the predicted upper bound of the two-phase BCC+B2 field. (b) Feature contributions to the predicted precipitate fraction at 1300 °C. Positive SHAP values increase the predicted target, whereas negative values decrease it; point color indicates the corresponding feature value.

The acquisition function is the weak element of the workflow. Scoring a fresh pool of 10,000 candidates with the weights of Section 2.4 produces a ranking that is anti-correlated with predicted stability, r = −0.29, and positively correlated with predicted phase count, r = +0.38, despite phase count carrying the largest negative weight in the score. The top 50 candidates average a predicted field of 1430 °C against 1756 °C for the unranked pool, so the ranking performed worse than no ranking at all. Figure 9 traces this to κ. Below about κ = 2 the score behaves as intended, and above it the exploration terms invert both signs. At the value used here, κ = 3, Hf- and Zr-bearing alloys fall from 50 of the top 50 to roughly 20, and the mean Al content of the top 50 rises from near zero to about 29 at.%. The Al-rich and Zr-rich drift reported in Section 3.5 is that effect rather than a discovery about the alloy system.

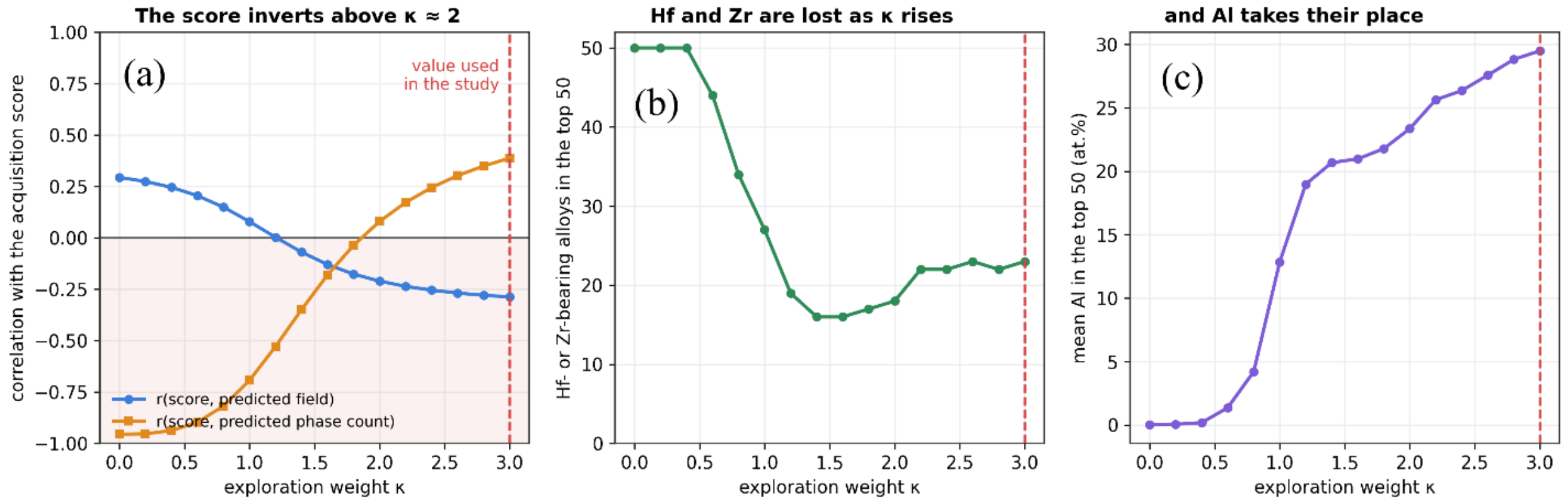

**Figure 9.** Behaviour of the acquisition function against the exploration weight κ. (A) Correlation of the score with predicted stability and with predicted phase count, both of which change sign above κ ≈ 2. (B) Hf- and Zr-bearing alloys among the 50 highest-scoring candidates. (C) Mean Al content of the same 50 candidates. The dashed line marks the value used in this work.

Set against the reference CALPHAD screen, the equilibrium results themselves hold up (Figure 10). Kube and co-workers predict that RuHf and RuZr systems remain stable to the solidus with no solvus transition, that RuTi does solutionize, and that RuAl solutionizes alongside V and Mo but forms sigma and A15 phases alongside Nb and Ta [3]. The present dataset reproduces all of it. No Hf-only or Zr-only alloy carries a solvus more than 50 °C below its solidus, 53% of Ti-only alloys do, sigma and A15 occur in none of the 255 Al-free alloys, and among 51 alloys carrying Ru with no group-IV or Al partner only 8% develop a two-phase field at all, with a median precipitate fraction of zero. Absolute temperatures sit 100 to 150 °C below the reference values, which is consistent with the use of TCHEA4 here against TCHEA5 there. The agreement across the whole Ru-B2 taxonomy, obtained from a different database version and a different sampling scheme, indicates that the thermodynamic layer of this workflow is sound and localizes the failure entirely in the screening layer above it.

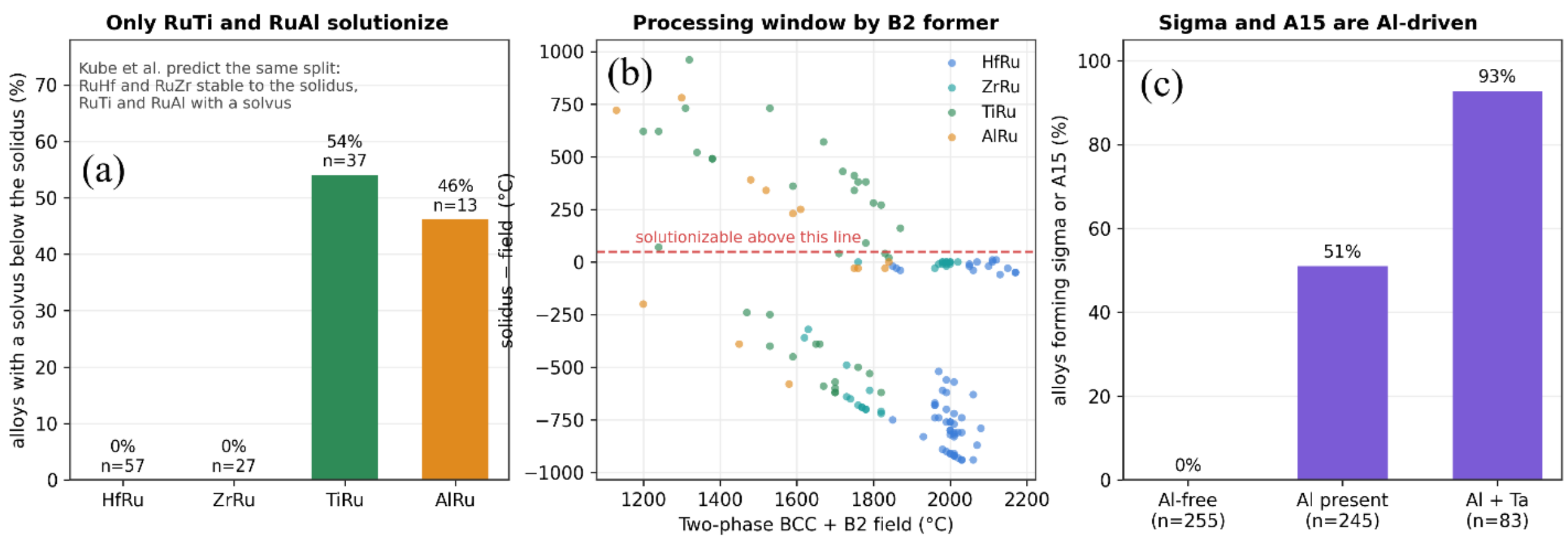

**Figure 10.** Agreement with the reference CALPHAD screen. (a) Fraction of alloys carrying a solvus below the solidus, by B2-forming addition. (b) Processing window against the two-phase field for each B2 former. (c) Incidence of sigma and A15 phases with and without Al.

## 4. Conclusions

This work establishes a machine-learning-guided CALPHAD framework for exploring multicomponent Ru-bearing BCC+B2 refractory alloys beyond the pseudobinary composition spaces considered previously. Equilibrium calculations for 500 compositions reproduce the principal thermodynamic trends established experimentally for Ru-B2 alloys. RuHf and RuZr provide exceptional high-temperature B2 stability, whereas RuTi more frequently provides a solutionizing window below the solidus. The calculations also recover the strong tendency for Al-containing compositions to form competing sigma and A15 phases. The agreement with established Ru-B2 behavior, including the recovery of two previously reported HfRu-B2 compositions within the successful alloy set, supports the use of the calculated dataset for broader compositional screening. Several practical alloy-design trends emerge from multicomponent search. Increasing Ru raises the upper bound of the two-phase BCC+B2 field rapidly up to approximately 9–10 at. % Ru, beyond which the benefit becomes comparatively small. Among the group-IV additions, Hf provides the greatest two-phase stability, followed by Zr and Ti. Ta contributes to high-temperature stability, while V is particularly effective in maintaining a phase-pure BCC+B2 constitution. In contrast, Al and Y account for most of the competing-phase formation observed in the present design space. These trends show that high B2 stability alone is insufficient and that useful refractory-superalloy design requires simultaneous control of the two-phase field, melting temperature, phase purity, and precipitate fraction. Applying these four physically based screening criteria identifies 100 alloys from the 500 evaluated compositions, of which 19 contain no more than 10 at.% Ru. The leading Ru-lean compositions are Nb28Mo23Ti19Ru9Hf11Zr10, Nb50Mo30Ru10Hf10, Nb52Ta28Ru9Hf11, and Nb55Ta20Ru10Hf15. The recovery of Hf-rich Nb–Mo, Nb–Ta, and Nb–V chemistries is consistent with the experimentally established stability of HfRu-B2 precipitates and indicates that the multicomponent interior of the Ru-B2 design space contains substantial opportunities beyond fixed pseudobinary composition paths.

The post-analysis of the machine-learning workflow also shows that the surrogate models themselves provide useful predictive accuracy, while the acquisition function strongly influences which regions of composition space are explored. At $\kappa = 3$, independently normalized uncertainty terms became large enough to bias the ranking toward uncertain compositions, and the use of total

BCC-B2 fraction did not distinguish a precipitation-strengthened two-phase alloy from a single-phase BCC-B2 state. Explicit inclusion of the precipitate fraction therefore provides an important physical constraint on the screening process. The compositional atomic-size mismatch descriptor remains useful as a low-cost proxy for lattice compatibility, but it should be interpreted as a screening descriptor rather than as the phase-specific BCC/B2 lattice misfit itself. The present study is limited to equilibrium phase stability and composition-level descriptors and does not explicitly resolve kinetic effects, mechanical properties, or phase-specific coherency. Future work will incorporate equilibrium phase compositions and phase-specific BCC/B2 lattice misfit into the screening framework and experimentally assess the most promising Ru-lean candidates.

## Data and code availability

The 500 Thermo-Calc step diagrams, the screening and random-forest code, and the scripts that generate every figure are available at https://github.com/mahata-lab/R2-Misfit-Calphad-ML. Every number reported here can be regenerated from the raw Thermo-Calc output.